\documentclass[aps,pra,twocolumn,showpacs,superscriptaddress,groupedaddress]{revtex4}
\usepackage{graphicx} 
\usepackage{amsmath}

\begin{document}

\title{One-dimensional quantum droplets under linear gravitational-like trap}

\author{Saurab Das}
\affiliation{Indian Institute of Information Technology Vadodara, Gujarat, India 382 028}
\author{Jayanta Bera}
\affiliation{C. V. Raman Global University, Bhubaneswar, Odisha 752 054, India}
\author{Ajay Nath}
\affiliation{Indian Institute of Information Technology Vadodara, Gujarat, India 382 028}


\begin{abstract}
We investigate the influence of a constant and time-dependent linear gravitational-like potential on one-dimensional (1D) quantum droplets (QDs), governed by an extended Gross–Pitaevskii equation (eGPE) incorporating a repulsive cubic effective mean-field (EMF) term and an attractive quadratic beyond-mean-field (BMF) correction. Within a tailored external confinement, we analytically  characterize the QDs wavefunction and derive the effective interaction contributions. Analogous to classical Newtonian dynamics, the falling velocity of the droplet within a finite domain is found to depend solely on the strength of the linear gravitational-like potential, remaining independent of both the total atom number and the magnitude of EMF nonlinearity. When the linear potential is temporally modulated, deviations in the trajectory of the droplet emerge relative to the static case, indicating potential applicability in precision gravimetry. To further probe the dynamical coherence properties, we compute the Shannon entropy and the Wigner quasi-probability distribution. Both measures reveal distinct signatures of the constant and time-varying linear potential, with the modulation strength directly influencing the phase-space localization and coherence structure of the droplet. Numerical simulations substantiate the stability of the analytical solutions, demonstrating their robustness. These findings suggest promising implications for quantum sensing and metrological applications using ultradilute quantum fluids.
\end{abstract}

\maketitle

\section{Introduction}
Ultracold atomic systems have emerged as a versatile platform for simulating gravitational and relativistic phenomena through the engineering of synthetic spacetimes and quantum gravitational analogs. Controlled Bose–Einstein condensates (BECs) enable laboratory emulation of curved spacetime scenarios, including analog black holes, cosmological expansion, and Hawking-like radiation \cite{Elliott, Aguilera, Zoest, Steinhauer, Bidel, Garay, Becker, Frye, Thompson, Eckel}. Their ultralow expansion energies and long coherence times make BECs ideally suited for precision matter-wave interferometry \cite{Ketterle}, facilitating long-duration free-fall interrogation \cite{Nandi}, inertial sensing \cite{Kovachy, Dimopoulos}, and studies at the quantum–gravity interface \cite{Lamporesi}. Microgravity experiments have enabled stringent tests of the Einstein equivalence principle \cite{Schlippert}, macroscopic quantum coherence \cite{Kovachy}, and hypothetical interactions with dark energy or dark matter \cite{Jaffe, El-Neaj}. These systems also advance quantum-enhanced satellite gravimetry \cite{Bidel} and inertial navigation technologies \cite{Barrett}. Although gravity is weak at the quantum scale, ultracold atoms and neutrons in Earth's gravitational field or engineered linear potentials allow emulation of key gravitational effects \cite{Abele, Pedram}. Prototypical systems such as the quantum bouncer, exhibiting discrete energy levels in gravitational confinement, offer precision tools for studying quantum dynamics in effective curved geometries \cite{Jenke, Poli}. Such capabilities establish ultracold atoms—particularly BECs—as a precise and tunable medium for probing foundational aspects of quantum gravity.

\begin{figure}
    \centering
    \includegraphics[width=\columnwidth]{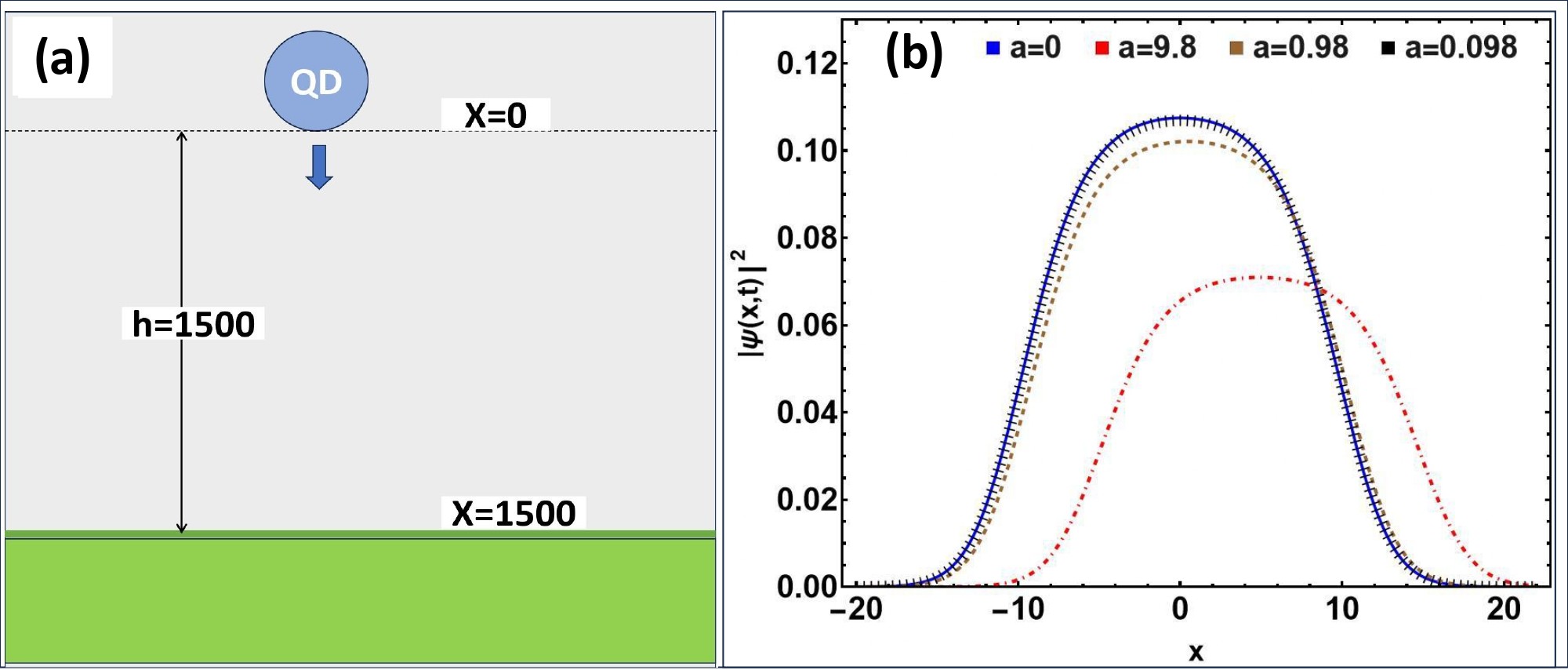}
    \caption{(a) Schematic representation of the theoretical model showing a QD released from an initial height of 1500 under free-fall conditions. (b) Spatial density profile of the QD as a function of position $x$ for different gravitational-like potential strengths $a$: $a = 0$ (free space, thick blue line), $a = 9.8$ (dot-dashed red line), $a = 0.98$ (dashed brown line), and $a = 0.098$ (dotted black line). The simulation is carried out at time $t = 1$, with parameters $\mu = \mu_0 = -\frac{2}{9}$, $G_1 = -1$, and $G_2 = 0.9999$. The spatial coordinate $x$ is scaled in units of the harmonic oscillator length. The figure illustrates the gravitational influence on QD density evolution and its sensitivity to linear potentials.}
    \label{fig1}
\end{figure}
In recent years, quantum droplets (QDs) have emerged as a robust class of self-bound ultracold matter stabilized by the interplay between attractive mean-field interactions and repulsive Lee–Huang–Yang (LHY) quantum fluctuations \cite{Malomed1,Luo,Petrov,Petrov1}. Initially observed in dipolar BECs of dysprosium and erbium atoms \cite{Ferrier,Schmitt,Hertkorn}, and later in binary contact-interacting BECs \cite{Cabrera,Cheiney,Semeghini}, QDs exhibit liquid-like coherence and structural stability even in ultradilute regimes—unlike bright solitons, which suffer from collapse and modulational instabilities. Their equilibrium and dynamics are well described by the extended Gross–Pitaevskii equation (eGPE), which incorporates the BMF effects \cite{Zin2018,Tylutki,Hu,Astrakharchik,Das,Maitri4}. External confinement crucially shapes QD behavior: harmonic traps influence size and collective modes \cite{Bhatia,Katsimiga,Guo,Bristy,Zezyulin,Fei,Decamps}, while optical lattices and anharmonic potentials induce spatial periodicities, symmetry breaking, and emergent nonlinear effects such as supersolidity \cite{Böttcher,Parit}, discrete time crystals \cite{Nath1}, vortex formation \cite{Dong2021}, soliton-droplet crossovers \cite{Pathak2022}, higher harmonics \cite{Maitri3}, and dark droplet states \cite{Edmonds2023}. Beyond static confinement, gravitational analogues offer a compelling platform to probe quantum dynamics under asymmetry and acceleration. QDs subjected to gravitational-like potentials exhibit interference, fragmentation, and Fermi-type acceleration \cite{Nandi,Benseghir2014,Siddik24}, while microgravity-like behavior can be engineered using linear optical potentials that simulate free-fall dynamics \cite{Shibata2020}. Additional studies have shown that linear and dimple potentials modify condensate properties such as atomic density and chemical potential \cite{Uncu2013,Naik2005}, providing enhanced control over droplet morphology and coherence. Motivated by these developments, the present study investigates the effects of constant and time-dependent linear gravitational-like potentials on 1D QDs. This choice is justified by their experimental relevance in terrestrial setups and the growing interest in using ultracold atoms to emulate quantum motion in curved or accelerating frames \cite{Kaltenbaek}, enabling a deeper understanding of how external asymmetries interact with self-binding and quantum fluctuations in ultradilute quantum liquids.

In this work, we develop an exact analytical model describing the formation and spatio-temporal evolution of QDs in a 1D mass-balanced binary BEC mixture subject to a linear gravitational-like potential. The system comprises a weakly interacting Bose–Bose mixture, where QD formation results from the interplay between repulsive intra-component and attractive inter-component interactions, as originally proposed by Petrov \cite{Petrov}. Within the framework of the 1D eGPE—incorporating a repulsive cubic EMF term and an attractive quadratic BMF correction—we derive closed-form analytical solutions for the QD wavefunction, phase profile, and effective nonlinearities using a similarity transformation technique. Our results explicitly capture the influence of the linear gravitational-like potential, revealing that the droplet’s falling velocity depends solely on the potential strength, independent of atom number and EMF nonlinearity—mirroring classical Newtonian motion. Temporal modulation of the linear potential produces measurable deviations in the droplet trajectory, indicating potential applications in precision gravimetry. To explore the system’s quantum coherence and excitation landscape, we compute the Wigner quasi-probability distribution in phase space, which uncovers intricate correlations between position and momentum beyond conventional density diagnostics. Additionally, Shannon entropy analysis highlights coherence modulation induced by the time-dependent linear field. Numerical simulations based on the split-step Fourier method corroborate the analytical findings, affirming the robustness and validity of the proposed theoretical model.

The remainder of the paper is organized as follows. In Sec. II, we present the theoretical model and derive closed-form analytical solutions of the 1D eGPE using a similarity transformation method, explicitly incorporating a linear gravitational-like potential. The resulting solutions yield the QDs wavefunction, phase profile, and effective nonlinearities, elucidating the role of the linear potential in governing droplet dynamics. In Sec. III, we analyze the spatio-temporal evolution of the droplet under static and time-dependent linear potentials, showing that the falling velocity depends solely on the potential strength—independent of atom number and BMF interactions. Sec. IV explores the system’s quantum coherence structure via the Wigner quasi-probability distribution in phase space, while Shannon entropy is employed to quantify coherence modulation induced by temporal variations in the potential. In Sec. V, we perform numerical simulations using the split-step Fourier method to validate the analytical predictions. Finally, Sec. VI summarizes the main results and outlines future perspectives.

\section{Analytical framework for droplet setting and linear trap} We consider a 1D homonuclear binary BEC composed of two equally populated and mass-balanced spinor components, i.e., $m_1 = m_2 \equiv m$ and $N_1 = N_2 \equiv N$, with symmetric intra-component repulsive interactions $g_{\uparrow\uparrow} = g_{\downarrow\downarrow} \equiv g > 0$, and inter-component attraction $g_{\uparrow\downarrow} < 0$ \cite{Petrov,Astrakharchik}. Under the condition $
\delta g = g_{\uparrow\downarrow} + \sqrt{g_{\uparrow\uparrow} g_{\downarrow\downarrow}} > 0,$ the system supports the formation of self-bound quantum droplets in 1D \cite{Petrov1}. This symmetry ensures identical equilibrium densities of both components, simplifying the analysis and enhancing the interpretability of the results. 

To investigate dynamical responses to external perturbations or simulate synthetic gravitational analogs, we introduce a time-dependent linear gravitational like potential along the axial direction.,
\begin{equation}\label{eq:QD1}
    V(x,t) = a(t)x,
\end{equation}

with strong transverse confinement ensuring effective 1D dynamics by freezing transverse excitations. Here, $a(t) \neq 0$ and can be tuned to represent static ($a(t) = a$) or oscillatory varying ($a(t) = a[1+ \alpha \cos(\omega t)]$), mimicking gravitational forces or inertial frame effects. The effective dynamics of the system are governed by the 1D eGPE \cite{Guo,Luo}:
\begin{equation}\label{eq:QD2}
   i \hbar \frac{\partial \Psi}{\partial t} = -\frac{\hbar^2}{2m} \frac{\partial^2 \Psi}{\partial x^2} + \frac{\delta g}{2} |\Psi|^2 \Psi - \frac{\sqrt{m}}{\pi \hbar} g^{3/2} |\Psi| \Psi + V(x,t) \Psi, 
\end{equation}

where the first nonlinear term corresponds to repulsive mean-field interactions and the second term represents attractive quantum fluctuations scaling as $|\Psi|\Psi$, crucial for stabilizing the droplet phase~\cite{Petrov, Astrakharchik}. The interaction strengths \( g \) and \( \delta g \) can be precisely tuned in experiments using Fano-Feshbach resonances \cite{Cheiney,Semeghini} or confinement-induced resonances \cite{Lim}. Experimentally, such binary BECs have been realized using two hyperfine states of $^{39}$K e.g., $|F=1, m_F=-1>$ and $|F=1, m_F=0>$, which have demonstrated quantum droplet formation in 3D and quasi-1D geometries \cite{Cabrera, Semeghini,Chomaz}.

To obtain a dimensionless form of the 1D eGPE, we rescale the relevant physical quantities by choosing characteristic units based on the transverse confinement frequency \( \omega_\perp \). The time, length, and wavefunction are rescaled as $
t \to \omega_\perp^{-1} t,\quad x \to \sqrt{\frac{\hbar}{m\omega_\perp}}\,x,\quad 
\psi \to \left( \frac{m\omega_\perp}{\hbar} \right)^{1/4} \psi,$
respectively. The external potential parameter \( a(t) \), which appears in the linear potential term \( a(t)x \), is expressed in units of \( \hbar \omega_\perp \). Under this scaling, the dimensionless extended Gross-Pitaevskii equation becomes:
\begin{equation}\label{eq:QD3}
 i\frac{\partial \psi}{\partial t} = -\frac{1}{2} \frac{\partial^2 \psi}{\partial x^2} + g_2 |\psi|^2 \psi - g_1 |\psi| \psi + a(t)\,x\,\psi,   
\end{equation}

where the dimensionless interaction parameters are given by: $
g_2 = \frac{\delta}{2} \left( \frac{m }{\omega_\perp^3 \hbar^3} \right)^{1/2}, \quad
g_1 = \frac{1}{\pi} \left( \frac{m g^{2}}{\omega_\perp \hbar^3} \right)^{3/4}$. Here, \( g_{2} \), and \( g_{1} \) are strengths of the cubic EMF and LHY correction of quadratic BMF interactions, respectively \cite{Bristy}. Now, in order to obtain analytical solutions of equation (\ref{eq:QD3}) in the presence of a time-dependent linear gravitational like potential and to investigate the spatio-temporal dynamics of QDs, we consider the following ansatz:
\begin{equation}\label{eq:QD4}
\psi(x,t) = \sqrt{B(t)}\, U[\eta(x,t)]\, e^{i\theta(x,t)},
\end{equation}
where \( B(t) \) is the time-dependent amplitude, and \( \eta(x,t) = x + \gamma(t) \) denotes a traveling coordinate of the condensate. Here, \( \gamma(t) \) is a positive, time-dependent modulation function, and  \( -\gamma(t) \) corresponds to the center-of-mass position of the excitation~\cite{Roy}. Furthermore, we assume the phase \( \theta(x,t) \) to have a quadratic form:
\begin{equation}\label{eq:QD5}
\theta(x,t) = -\gamma'(t) x - \frac{1}{2} \int [\gamma'(t)^{2}+E] dt.
\end{equation}

Now, substituting  the ansatz equation (\ref{eq:QD4}) and phase form (\ref{eq:QD5}) into equation (\ref{eq:QD3}), we obtain the following consistency conditions:
\begin{equation}\label{eq:QD6}
-\frac{\partial^2 U}{\partial \eta^2}-G_1\mid F(\eta)\mid F(\eta) +G_2\mid F(\eta)\mid^{2} F(\eta)=E F(\eta),
\end{equation}
where $E$ is the eigenvalue of equation (\ref{eq:QD6}), $G_1$, $G_2$ denote the nonlinearity constants which can have positive or negative depending on scattering length along with:
\begin{equation}\label{eq:QD7}
B(t)=1, \; \; G_{2} =2 g_{2},\;\; 2 g_{1}-G_{1} =0 \;\; a(t) = \gamma''(t).
\end{equation}

The solution of equation (\ref{eq:QD6}) can be given as \cite{Petrov,Astrakharchik}:
\begin{equation}\label{eq:QD8}
		U[\eta]=\frac{ 3 \mu  }{1+\sqrt{1-\frac{\mu}{\mu_{0}} \frac{ G_{2}}{ G_{1}}  } \cosh (\sqrt{\text{-E}} \eta)}
		\end{equation}
where, $\mu=E/G_{1}$, $\mu_{0}=-2/9$ with $ E<0$, $G_1<0$ and $G_2>0$. Thus, the complete solution of equation (\ref{eq:QD3}) can be written as: 
\begin{equation}\label{eq:QD9}
\psi(x,t)=\frac{3 \mu e^{i\left[-\gamma'(t) x - \frac{1}{2} \gamma'(t)^{2} t \right]} }{1+\sqrt{1-\frac{\mu}{\mu_{0}} \frac{ G_{2}}{ G_{1}}  } \cosh (\sqrt{\text{-E}} ( x + \gamma(t)))}.
\end{equation}

For $\gamma(t) =0$, the external trap $V(x,t)=0$ i.e. free space and equation (\ref{eq:QD9}) reduce to QDs in free space \cite{Astrakharchik}. Using the above solution and the normalization condition: 
$N = \int_{-\infty}^{+\infty} |\psi(x,t)|^2 \, dx,$ the explicit relation between $N$ and $\mu$ for the case $G_1 > 0$ and $G_2 = -1$ can be obtained as:
\begin{equation}\label{eq:QD10}
N = \frac{4}{3} \left[ \ln \left( \frac{1 + \sqrt{\frac{\mu}{\mu_0}}}{\sqrt{1 - \frac{\mu}{\mu_0}}} \right) - \sqrt{\frac{\mu}{\mu_0}} \right].
\end{equation}
Remarkably, the expression in equation~(\ref{eq:QD10}) coincides with the result derived for quantum droplets in free space~\cite{Astrakharchik}, indicating that the presence of an external harmonic trap does not affect the conserved quantity $N$. This invariance reflects an underlying continuous symmetry of the system, consistent with Noether’s theorem~\cite{Sardanashvily} and consequently, despite external modulation, normalization $N$ remains a constant of motion. Using equation~(\ref{eq:QD9}), we now examine the spatio-temporal dynamics of quantum droplets under various experimentally realizable temporal modulations of the trapping frequency $a(t)$, by selecting analytically tractable forms of the modulation function $\gamma(t)$.

\section{Droplet dynamics under linear gravitational like trap} We analyze the dynamics of 1D QDs governed by the exact solution of the 1D eGPE [equation ~(\ref{eq:QD9})], incorporating temporally varying repulsive cubic EMF interactions and attractive quadratic BMF corrections in the presence of a linear gravitational-like potential. Utilizing the formalism established in equation ~(\ref{eq:QD7}), we explore the droplet evolution under experimentally feasible temporal modulations of the confinement strength $a(t)$, realized through appropriate choices of the phase function $\gamma(t)$, as follows:

\begin{itemize}
    \item[(a)] \textbf{ Free space:} $\gamma(t) = \text{constant},\quad a(t) = 0$;
    \item[(b)] \textbf{ Time-independent linear trap:} $\gamma(t) = -\frac{1}{2} a t^2,\quad a(t) = -a$;
    \item[(c)] \textbf{ Time-dependent linear trap:} $\gamma(t) = -\frac{a t^2}{2} + \frac{a \alpha \cos(\omega t)}{\omega^2},\quad a(t) = -a(1 + \alpha \cos(\omega t))$.
\end{itemize}

\begin{figure}
    \centering
    \includegraphics[width=\columnwidth]{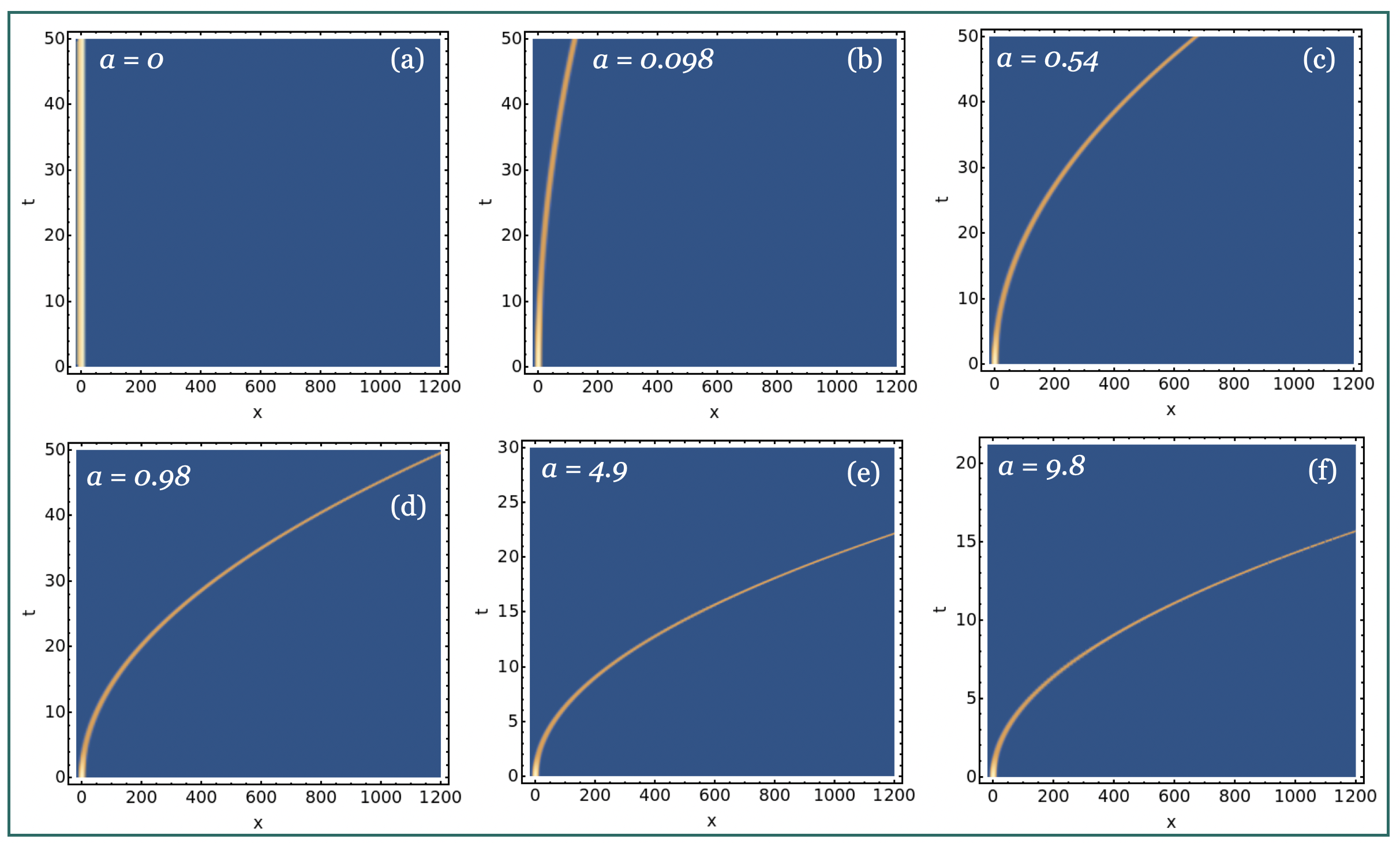}
    \caption{Temporal evolution of the QDs density at various heights under different gravitational-like acceleration strengths $a$. Subfigures (a)–(f) correspond to $a = 0$, $0.098$, $0.54$, $0.98$, $4.9$, and $9.8$ respectively. The results illustrate the effect of gravitational strength on the dynamical response of the QD, including increase in velocity profiles. The simulations are carried out for the parameters $\mu = \mu_0 = -\frac{2}{9}$, $G_1 = -1$, and $G_2 = 0.9999$. The spatial coordinate is normalized by the harmonic oscillator length.}
    \label{fig2}
\end{figure}

Here, $a$, $\alpha$, and $\omega$ are real, positive constants. This formulation enables tunable control over the linear confinement, thereby modulating droplet dynamics. Our results demonstrate that in a static linear potential, the droplet exhibits a constant falling velocity determined solely by the potential strength, independent of the atom number or BMF nonlinearity reminiscent of classical free-fall under Newtonian gravity. However, under temporal modulation of the linear potential, the trajectory deviates from uniform acceleration, manifesting sensitivity to the driving parameters. 

\begin{figure}
    \centering
    \includegraphics[width=\columnwidth]{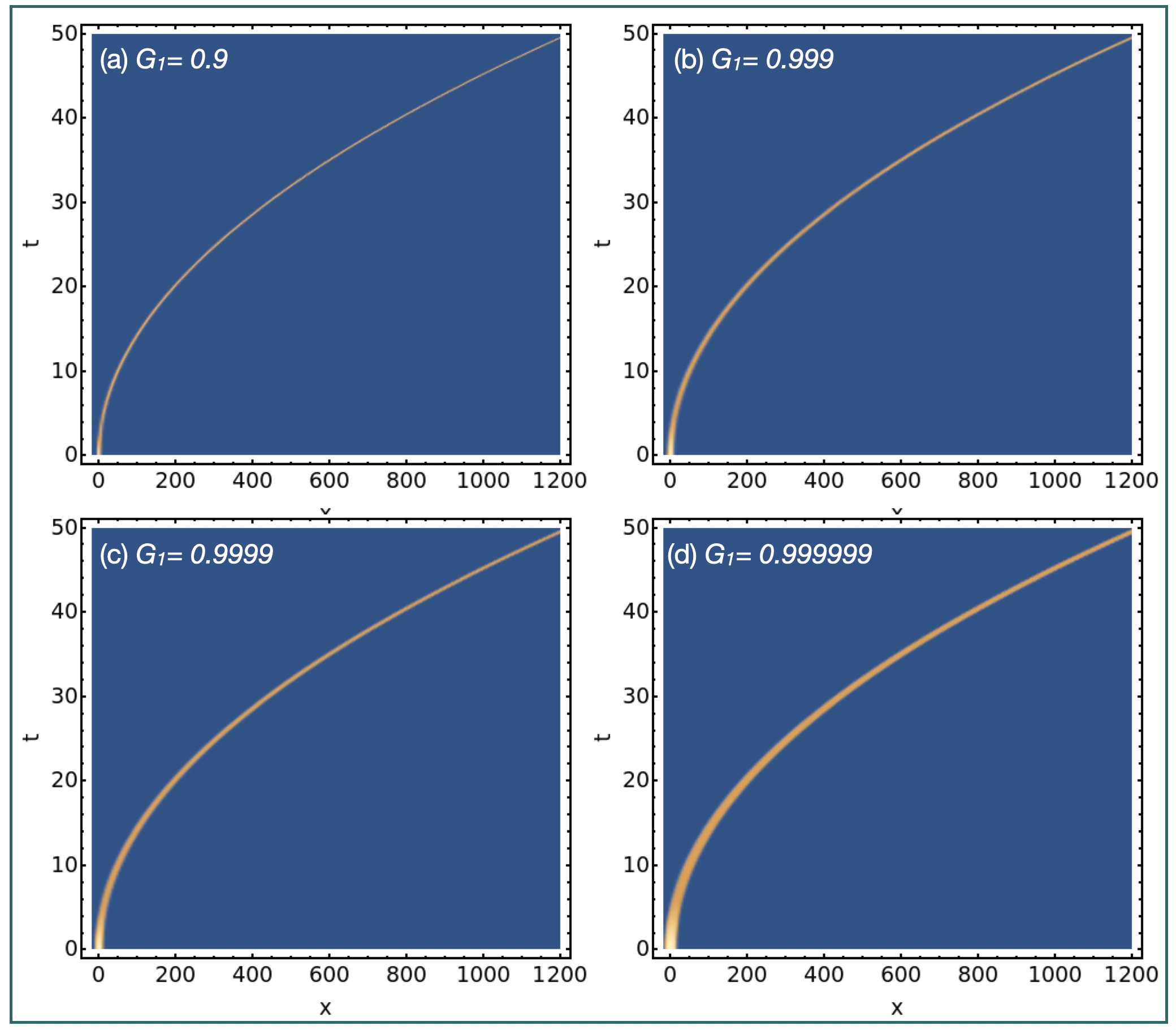}
    \caption{Temporal evolution of QDs density profiles at various heights for different strengths of the EMF interaction coefficient $G_{2}$, under a fixed gravitational-like acceleration $a = 0.098$. Subfigures (a)–(d) correspond to $G_{2} = 0.9$, $0.999$, $0.9999$, and $0.999999$, respectively. The simulations reveal the impact of beyond-mean-field (BMF) contributions on the QD dynamics, demonstrating enhanced density localization with increasing $G_{2}$. The parameters used are $\mu = \mu_0 = -\frac{2}{9}$ and $G_1 = -1$. The spatial coordinate is scaled by the harmonic oscillator length. These results highlight the role of EMF-BMF interplay in shaping droplet structure and stability.}
    \label{fig3}
\end{figure}

\subsection{Time-Independent linear gravitational like trap} In order to investigate the dynamics of QDs under free fall scenario i.e. under constant acceleration, we consider $\gamma(t) =- (1/2) a t^2$ resulting into the form of constant linear trap configuration $V(x,t) =- a x$ along with EMF and BMF strengths are $G_{2} =2 g_{2},\;\; G_{1}=2 g_{1}$, respectively. Here, $a$ is real constant akin to acceleration due to gravity. For the chosen form of $\gamma(t)$, the complete form of wavefunction from equation (\ref{eq:QD9}) becomes: 

\begin{equation}
    \psi(x,t)=\frac{\frac{3\mu}{G_1}\times e^{i[-atx+\alpha(t)]}}{1+\sqrt{1-\frac{\mu}{\mu_0}\frac{G_2}{G_1^2}}cosh[\sqrt{-\mu}(x-\frac{a t^2}{2})]}, 
\end{equation}
 
with $\alpha(t) =-(1/2)[a^2 t^3-E t ]$. Figure~\ref{fig1}(a) illustrates the classical free-fall trajectories of QDs released from the different heights under a tunable gravitational-like potential. The corresponding spatiotemporal evolution of the QD density is shown in Figure~\ref{fig1}(b) for increasing gravitational acceleration \( a \). In the absence of gravity (\( a = 0 \)), the QD remains centered around \( x = 0 \) with maximal amplitude, representing a stable, self-bound state. As the gravitational strength increases (e.g., \( a = 9.8 \)), the density peak shifts toward positive \( x \), accompanied by a reduction in central amplitude and broadening of the spatial profile—indicative of enhanced gravitational spreading and redistribution of mass. When \( a \) is reduced to intermediate values (e.g., \( a = 0.98 \) or \( 0.098 \)), the density progressively re-localizes around the origin, recovering its initial peak structure. This reversible transition highlights how gravity modulates the internal configuration and localization properties of the QD without disrupting its integrity. All simulations are performed at a fixed time \( t = 1 \) with parameters \( \mu = -2/9 \), \( G_1 = -1 \), and \( G_2 = 0.9999 \), using dimensionless oscillator units.

To further examine the dynamical response, figure~(\ref{fig2}) displays the full temporal and spatiotemporal evolution of the QD density for a range of gravitational accelerations \( a = 0 \), \( 0.098 \), \( 0.54 \), \( 0.98 \), \( 4.9 \), and \( 9.8 \) [Figures~\ref{fig2}(a)-\ref{fig2}(h)]. As \( a \) increases, the density maxima exhibit a consistent rightward drift along both space and time axes, indicating an increase in center-of-mass velocity under stronger gravitational pull. Remarkably, throughout this motion, the QD retains its sharply localized internal structure—characterized by persistent density spikes—despite the continuous external drive. This robustness underscores the self-bound nature of QDs and their resilience to linear acceleration, where gravitational forces alter translational motion without significantly deforming internal coherence. These simulations use the same parameter set as before, with spatial and temporal coordinates scaled by the transverse harmonic oscillator length and frequency, respectively.

Figure~(\ref{fig3}) illustrates the temporal evolution of QD density profiles at different heights for varying values of the EMF interaction coefficient \( G_{2} \), while keeping the gravitational-like acceleration fixed at \( a = 0.098 \). Subfigures~\ref{fig3}(a)--\ref{fig3}(d) correspond to \( G_{2} = 0.9 \), \( 0.999 \), \( 0.9999 \), and \( 0.999999 \), respectively. As the value of \( G_{2} \rightarrow 1\), the density profiles become increasingly sharper and more localized, indicating stronger self-binding of the QD. Notably, since the gravitational acceleration \( a \) remains constant across all cases, no change is observed in the velocity of the QD’s fall, consistent with classical Newtonian dynamics where acceleration—and hence velocity change—is independent of mass or internal structure. These results emphasize the critical role of EMF contributions in modulating the internal coherence of the QD, while the overall translational motion adheres to classical expectations.

\begin{figure}
    \centering
    \includegraphics[width=\columnwidth]{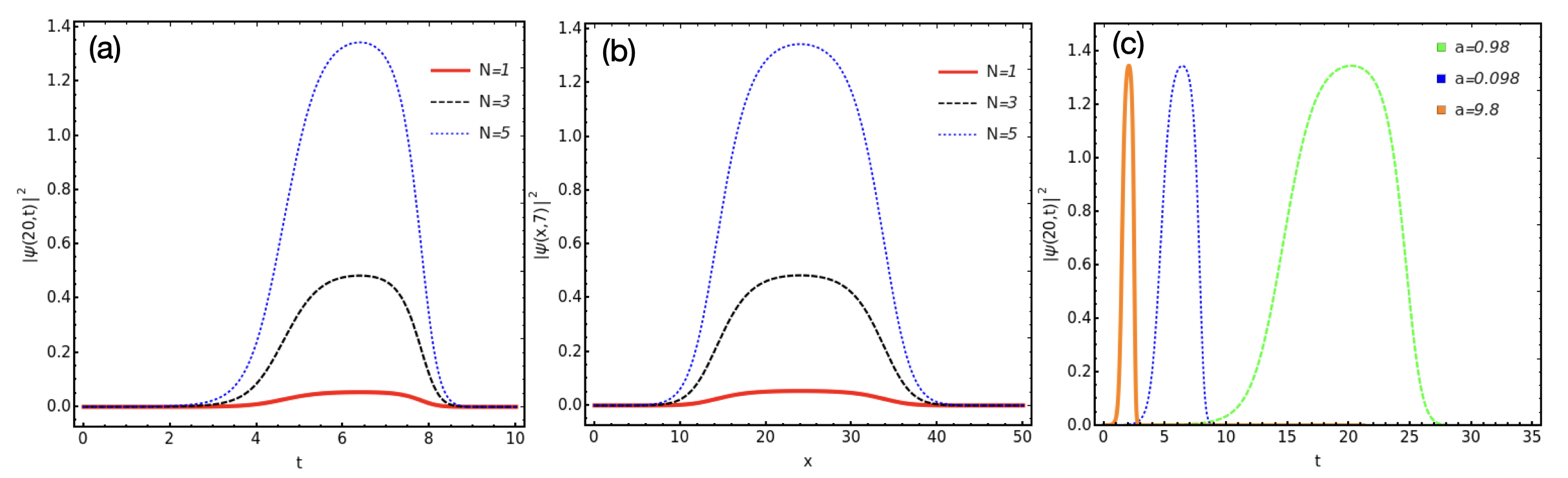}
    \caption{Dynamics of QDs density profiles for varying atom numbers \( N \) and gravitational-like strength \( a \). (a) Temporal evolution of the density \( |\psi(x,t)|^2 \) at a fixed position \( x = 20 \) for \( N = 1, 3, 5 \) at \( a = 0.98 \); (b) Spatial profile of the density at time \( t = 7 \) for the same values of \( N \); (c) Temporal evolution at \( x = 20 \) for fixed \( N = 5 \) and different values of gravitational strength \( a = 9.8,\, 0.98,\, 0.098 \). Parameters used: \( \mu = \mu_0 = -\frac{2}{9} \), \( G_1 = -1 \), \( G_2 = 0.9999 \), with spatial coordinates scaled by the harmonic oscillator length.}
    \label{fig4}
\end{figure}

In figure (\ref{fig4}), the temporal and spatial evolution of QDs density profiles is illustrated under varying atom numbers \( N = 1, 3, 5 \), while subject to a fixed gravitational-like acceleration. In figure \ref{fig4}(a), the density \( |\psi(x,t)|^2 \) is plotted at a fixed position \( x = 20 \) as a function of time. It is observed that, for all three cases, the QDs remain localized at \( x = 20 \) from approximately \( t = 2 \) to \( t = 8.5 \), indicating that the velocity of fall is independent of the number of atoms - a result consistent with Newtonian physics where gravitational acceleration is mass-independent. However, the magnitude of the density increases with increasing \( N \), reflecting improved internal interactions that lead to stronger self-binding. Figure \ref{fig4}(b) presents the spatial density profile at a fixed time \( t = 7 \), showing increasingly sharper and higher peaks as \( N \) increases, further confirming the role of the interatomic interaction strength in shaping the internal structure of the QD. In figure \ref{fig4}(c), for a fixed atom number \( N = 5 \), the density \( |\psi(x,t)|^2 \) is again plotted at \( x = 20 \) as a function of time for different gravitational strengths \( a = 0.098, 0.98, 9.8 \). As expected, with decreasing \( a \), the arrival time of the density peak at \( x = 20 \) is delayed, which is consistent with previous observations, affirming that the falling dynamics are governed by classical acceleration, while the internal density structure remains dictated by quantum interactions.

\begin{figure}
    \centering
    \includegraphics[width=\columnwidth]{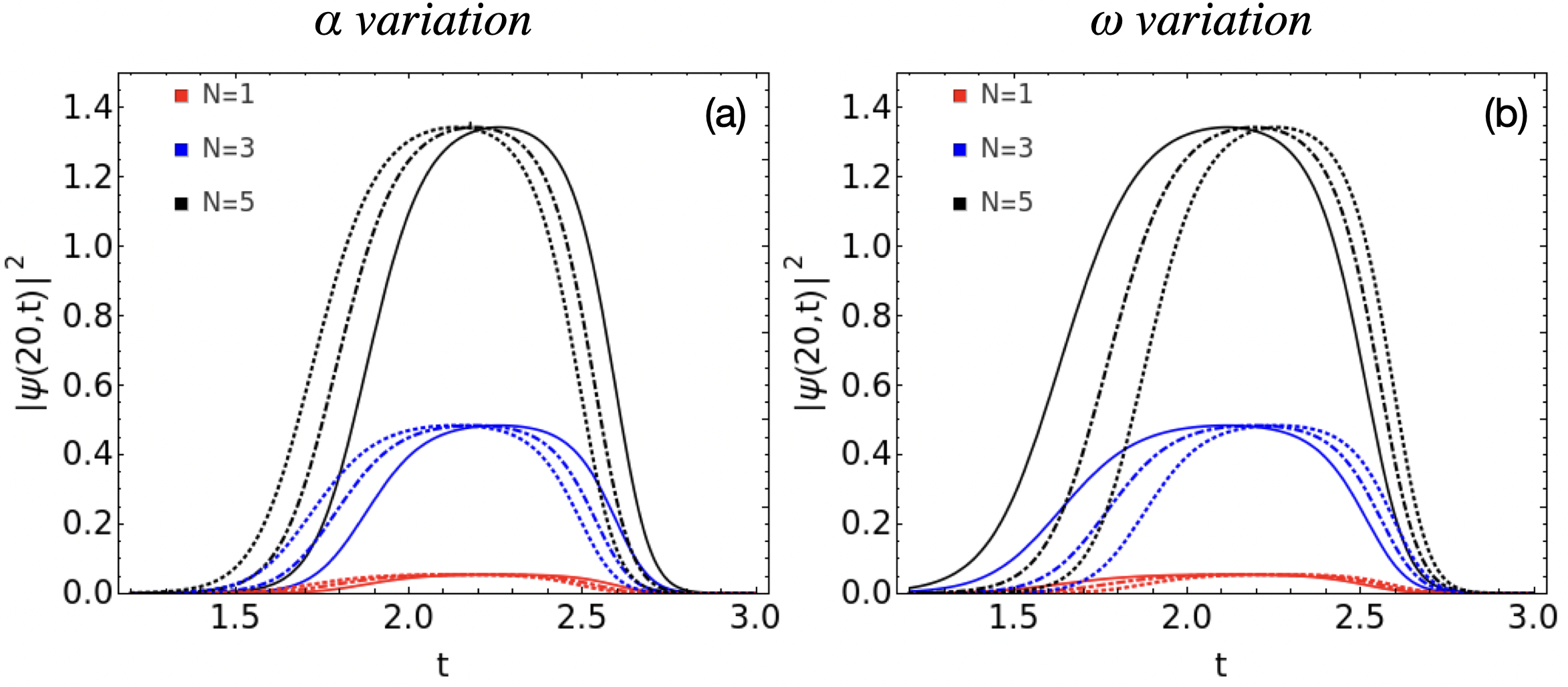}
    \caption{Temporal dynamics of QDs density profiles under a time-dependent gravitational-like potential of the form \( -a \left(1 + \alpha \cos(\omega t)\right) \). (a) Time evolution of the density \( |\psi(x,t)|^2 \) at a fixed position \( x = 20 \) for \( N = 1, 3, 5 \), with \( a = 0.98 \), \( \omega = 0.5 \), and modulation amplitudes \( \alpha = 0.1 \) (red), \( \alpha = 0.2 \) (blue), and \( \alpha = 0.3 \) (black). (b) Time evolution of the density \( |\psi(x,t)|^2 \) at \( x = 20 \) for \( N = 1, 3, 5 \), with \( a = 9.8 \), \( \alpha = 0.3 \), and driving frequencies \( \omega = 0.5 \) (red), \( \omega = 0.55 \) (blue), and \( \omega = 0.60 \) (black). The chosen physical parameters: \( \mu = \mu_0 = -\frac{2}{9} \), \( G_1 = -1 \), \( G_2 = 0.9999 \); spatial coordinates are scaled by the harmonic oscillator length. }
    \label{fig5}
\end{figure}

\subsection{Time-dependent linear gravitational like trap} Next, we investigate the dynamics of QDs under time-varying gravitational-like potentials during free fall, providing a platform to probe the dynamical response of dilute, self-bound quantum fluids to non-inertial and time-dependent external fields, enriching our understanding of quantum hydrodynamics beyond the mean-field regime. This line of inquiry offers deeper insights into gravitational analogs in ultracold systems, potentially contributing to the development of precision quantum sensors, quantum thermometric techniques, and analog models of curved spacetime in BECs. For realizing a time-dependent gravitational like potential, we consider $\gamma(t)= -\frac{a t^2}{2} + \frac{a \alpha \cos( \omega t)}{\omega^2}$ resulting in the trap form: $V(x,t) = -a \left(1 + \alpha \cos(\omega t)\right)) x$ from equation (\ref{eq:QD7}). The corresponding wavefunction and phase forms can be estimated from equation (\ref{eq:QD9}) and equation (\ref{eq:QD5}), respectively.

The temporal dynamics of QDs under a time-dependent gravitational-like potential of the form $-a(1 + \alpha \cos(\omega t))$ exhibits a notable sensitivity to variations in both the modulation amplitude $\alpha$ and the driving frequency $\omega$. Figures~\ref{fig5}(a) and (b) show the evolution of the QD density $|\psi(x,t)|^2$ at a fixed spatial position $x = 20$ for different number of particles $N = 1, 3, 5$, with a fixed potential strength $a = 9.8$. In Fig.~\ref{fig5}(a), with constant driving frequency $\omega = 0.5$, increasing the modulation amplitude from $\alpha = 0.1$ (red) to $\alpha = 0.2$ (blue) and $\alpha = 0.3$ (black) leads to an earlier arrival of the droplet and sharper peaks in the local density profile, indicating enhanced gravitational pull with increasing $\alpha$. Similarly, in figure~\ref{fig5}(b), with $\alpha = 0.3$ fixed, increasing $\omega$ from $0.5$ (red) to $0.55$ (blue) and $0.6$ (black) causes a noticeable shift in the temporal location of peak densities, highlighting the sensitivity of QD motion to external driving frequency. These effects become increasingly prominent with higher $N$, as the peak density and dynamical responsiveness increase due to the enhanced nonlinearity. Here, $G_1 = -1$, $G_2 = 0.9999$, and the chemical potential is fixed at $\mu = \mu_0 = -2/9$. The variation of local density with $\alpha$ and $\omega$ is consistent with previous theoretical studies on modulated nonlinearity and parametric resonance in quantum fluids~\cite{mistakidis2022cold}. The results reveal a tunable mechanism for manipulating QD transport and density location via time-periodic gravitational analogs. 

\begin{figure}
    \centering
    \includegraphics[width=\columnwidth]{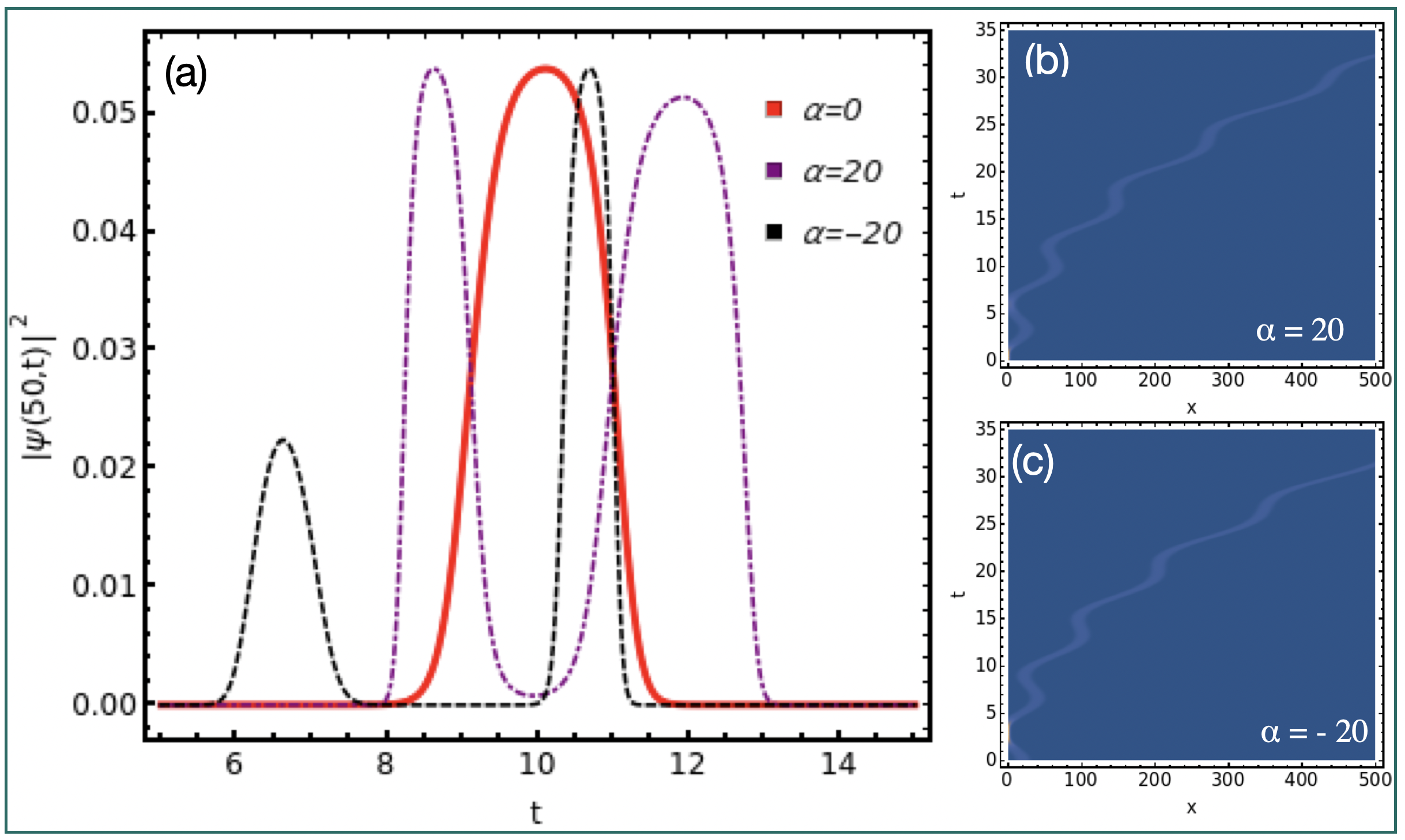}
    \caption{Temporal dynamics of QDs density profiles under a time-dependent gravitational-like potential of the form \( -a \left(1 + \alpha \cos(\omega t)\right) \). (a) Time evolution of the density \( |\psi(x,t)|^2 \) at a fixed position \( x = 50 \) for \( a = 0.98 \), \( \omega = 1 \), and modulation amplitudes \( \alpha = 0 \) (red), \( \alpha = 20 \) (dot-dashed purple), and \( \alpha = -20 \) (dashed black). Temporal evolution of QDs density profiles at various heights for different strengths of the (b) $\alpha = 20$, \& (c) $\alpha = -20$, under a fixed $a = 0.098$. The chosen physical parameters: \( \mu = \mu_0 = -\frac{2}{9} \), \( G_1 = -1 \), \( G_2 = 0.9999 \); spatial coordinates are scaled by the harmonic oscillator length. }
    \label{fig6}
\end{figure}

Figure (\ref{fig6}) shows the temporal dynamics of QDs density profiles under the influence of a time-dependent gravitational-like potential of the form $-a \left(1 + \alpha \cos(\omega t)\right)$. Panel (\ref{fig6})(a) presents the time evolution of the density $|\psi(x,t)|^2$ at a fixed spatial position $x = 50$, for three modulation amplitudes: $\alpha = 0$ (red), $\alpha = 20$ (dot-dashed purple), and $\alpha = -20$ (dashed black), under the fixed parameters $a = 0.98$, $\omega = 1$. For the unmodulated case $\alpha = 0$, the density reaches $x = 50$ at time $t = 8.5$, and no significant oscillatory motion is observed afterward, indicating propagation at constant linear velocity. In contrast, for $\alpha = 20$, the density peak reaches the same location earlier, at $t = 8$, and is observed to oscillate back to $x = 50$ at $t = 10.5$, suggesting the generation of acceleration due to the time-periodic modulation. Similarly, for $\alpha = -20$, the peak arrives much earlier at $t = 6$, and again revisits the position at $t = 10$, further highlighting the impact of modulation phase on induced dynamics. Panels (b) and (c) provide the full space-time evolution of $|\psi(x,t)|^2$ for $\alpha = 20$ and $\alpha = -20$, respectively. These results clearly illustrate the transition from constant velocity transport (as seen in earlier static potential cases) to accelerated dynamics due to time-varying potentials. The modulation introduces a periodically varying force, resulting in temporal changes in acceleration and effective group velocity of the droplet. The asymmetry in the behavior for positive and negative $\alpha$ values emphasizes the directional dependence introduced by the phase of the oscillating field. The observed dynamics are robust across different modulation amplitudes and provide insights into force engineering using temporal modulations in quantum fluids. The simulations were performed for a chemical potential $\mu = \mu_0 = -\frac{2}{9}$, with nonlinear coefficients $G_1 = -1$ and $G_2 = 0.9999$.

\section{Phase-Space and Information-Theoretic study of falling Droplets}

\begin{figure}
    \centering
    \includegraphics[width=1.0\linewidth]{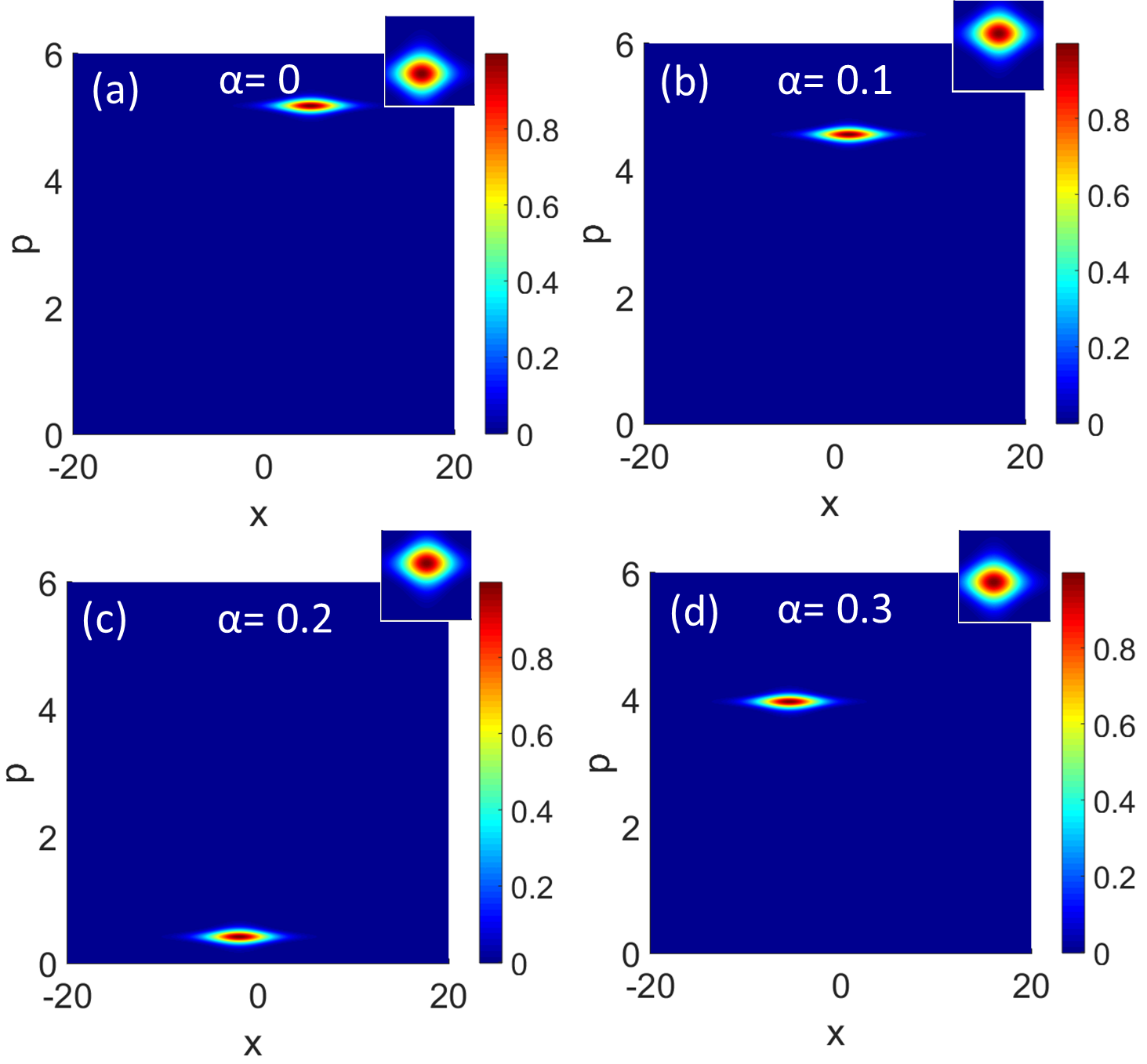}
    \caption{Wigner quasi-probability distribution \( W(x, p) \) illustrating the phase-space dynamics of QDs at time \( t = 1 \), under trap form \( V(x,t) = -a(1 + \alpha \cos(\omega t)) x \) for (a) \( \alpha = 0 \), (b) \( 0.1 \), (c) \( 0.2 \), and (d) \( 0.3 \), respectively, with \( a = 9.8 \), \( \omega = 0.5 \), \( \mu = \mu_0 = -\frac{2}{9} \), \( G_1 = -1 \), and \( G_2 = 0.9999 \). The spatial coordinates are scaled by the harmonic oscillator units. }\label{fig7}
\end{figure}

\begin{figure}
    \centering
    \includegraphics[width=1.0\linewidth]{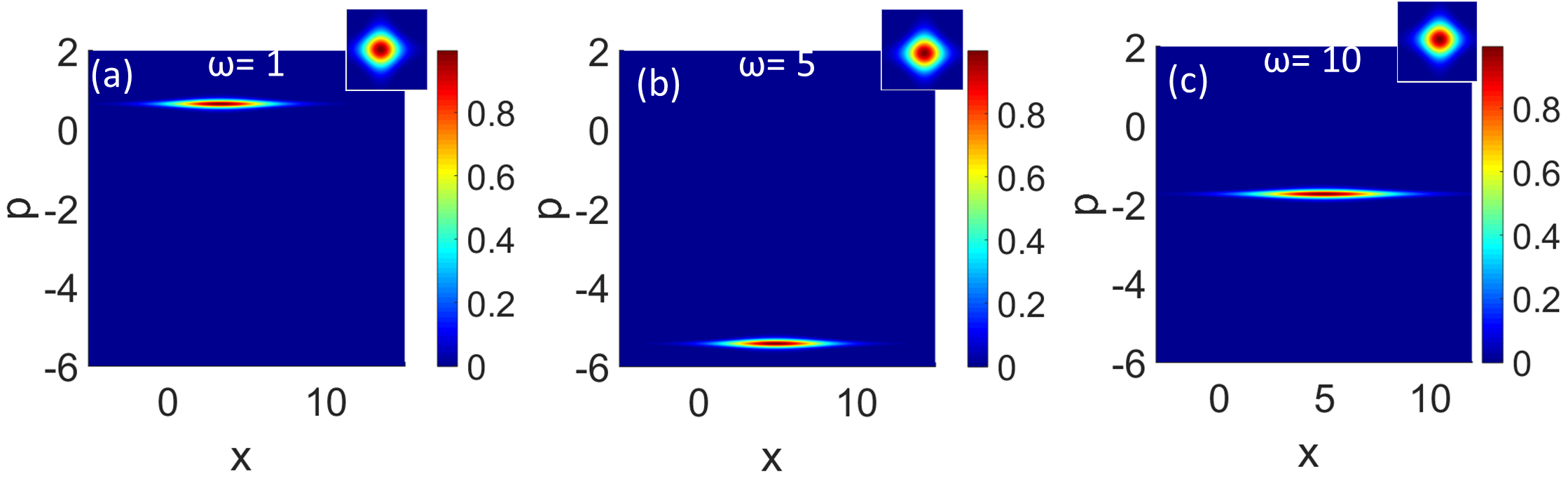}
    \caption{Wigner quasi-probability distribution \( W(x, p) \) illustrating the phase-space dynamics of QDs at time \( t = 1 \), under trap form \( V(x,t) = -a(1 + \alpha \cos(\omega t)) x \) for (a) \( \omega = 1 \), (b) \( 5 \), and (c) \( 10 \), respectively, with \( a = 9.8 \), \( \alpha = 0.3 \), \( \mu = \mu_0 = -\frac{2}{9} \), \( G_1 = -1 \), and \( G_2 = 0.9999 \). The spatial coordinates are scaled by the harmonic oscillator units.}\label{fig8}
\end{figure}

\subsection{Wigner Phase-Space distribution}
In the dynamical analysis of 1D QDs) falling under linear gravitational-like potentials, characterizing the internal coherence and excitation properties is crucial for understanding the interplay between quantum self-binding and external driving forces. While the evolution of the density profile offers information about spatial deformation or location, these metrics alone are insufficient to reveal deeper correlations between position and momentum—essential for capturing phase coherence and identifying non-classical excitation modes within the droplet. To address this, we employ the Wigner quasi-probability distribution function as a diagnostic tool to visualize the droplet's phase-space dynamics \cite{Wigner,Verstraete}. This representation enables a unified description of quantum states in terms of both position and momentum, making it especially suitable for analyzing QDs under gravitational-like acceleration where time-dependent modulation influences both spatial and momentum degrees of freedom. However, these are quasi-probability distributions, not true probabilities, due to the non-classical nature of the Wigner function. The Wigner function is given by \cite{Zurek,Vitali}:

\begin{equation}
W(x,p) = \frac{1}{\pi\hbar} \int_{-\infty}^{\infty} \psi^*(x + x_0, t) \psi(x - x_0, t) \, e^{2 i x_0 p / \hbar} \, dx_0.
\end{equation}

where $\psi(x,t)$ is the droplet wavefunction obtained from the equation (\ref{eq:QD9}) for chosen $\gamma(t)$. The Wigner function provides a fine-grained visualization of the phase-space structure and is particularly sensitive to quantum interference phenomena, which may manifest during the droplet's free fall or in response to periodic modulations of the gravitational-like potential.

To analyze the phase-space dynamics of QDs under time-modulated linear gravitational-like confinement, we evaluate the Wigner quasi-probability distribution function \( W(x, p) \) at time \( t = 1 \), using the potential \( V(x,t) = -a(1 + \alpha \cos(\omega t))x \). Figures~\ref{fig7}(a)–(d) present the evolution of \( W(x, p) \) for modulation amplitudes \( \alpha = 0, 0.1, 0.2 \), and \( 0.3 \), respectively, with fixed parameters \( a = 9.8 \), \( \omega = 0.5 \), \( \mu = \mu_0 = -\frac{2}{9} \), \( G_1 = -1 \), and \( G_2 = 0.9999 \). The distribution exhibits characteristics akin to coherent states, manifesting as localized Gaussian-like lobes in phase space with varying widths and positions for different values of \( \alpha \). As the modulation amplitude increases, the center of mass of the distribution shifts, and its spread changes, indicating alterations in the droplet's coherence and momentum distribution. Notably, for \( \alpha = 0.2 \), the Wigner distribution shows the lowest average momentum, suggesting reduced dynamical excitation. To further investigate the frequency-dependent effects, we plot \( W(x, p) \) in figure~(\ref{fig8}) for fixed \( \alpha = 0.3 \) and varying driving frequencies \( \omega = 1, 5, 10 \). The inset shows a zoomed-out view over a \( 12 \times 0.5 \) phase-space region, revealing that the peak of the Wigner distribution migrates from positive to negative momentum values as \( \omega \) increases. This frequency-dependent asymmetry highlights the emergence of tunable momentum control and dynamical symmetry breaking in QD systems under time-periodic gravitational-like potentials, offering potential applications in momentum-resolved quantum sensing and interferometry.

\subsection{Estimation of Shanon Entropy} In the study of QD dynamics subject to linear gravitational-like confinement, information-theoretic measures offer critical insight into nonlocal features such as coherence, localization, and complexity of the underlying quantum state. Among these, the Shannon entropy—introduced by C. Shannon as the foundational quantity in information theory—has emerged as a versatile tool for characterizing the global spatial structure of quantum systems \cite{Shannon}. For a 1D QD described by the wavefunction \( \psi(x,t) \), the temporal evolution under external gravitational-like potentials modifies the spatial probability distribution \( |\psi(x,t)|^2 \). While standard observables such as the peak density track local properties, they fail to capture global delocalization or spreading behavior of the quantum state. To quantify such behavior, we evaluate the Shannon entropy in position space, defined as:
\begin{equation}
S_\rho(t) = - \int_{-\infty}^{\infty} |\psi(x,t)|^2 \log \left[|\psi(x,t)|^2\right] dx.
\end{equation}

Here, \( S_\rho(t) \) serves as a global indicator of informational content and localization: higher entropy implies greater spatial spread, while lower entropy indicates more localized and coherent structures. In the case of gravitational-like potentials of the form \( V(x,t) = -a(1 + \alpha \cos(\omega t))x \), the evolution of \( S_\rho(t) \) reveals how modulation in external driving alters the coherence and internal structure of the QD \cite{Siddik24}. Such analyses provide complementary insight to phase-space methods such as the Wigner function, and are particularly relevant for studying non-equilibrium and self-bound quantum fluids under time-dependent forcing.

\begin{figure}
    \centering
    \includegraphics[width=1.0\linewidth]{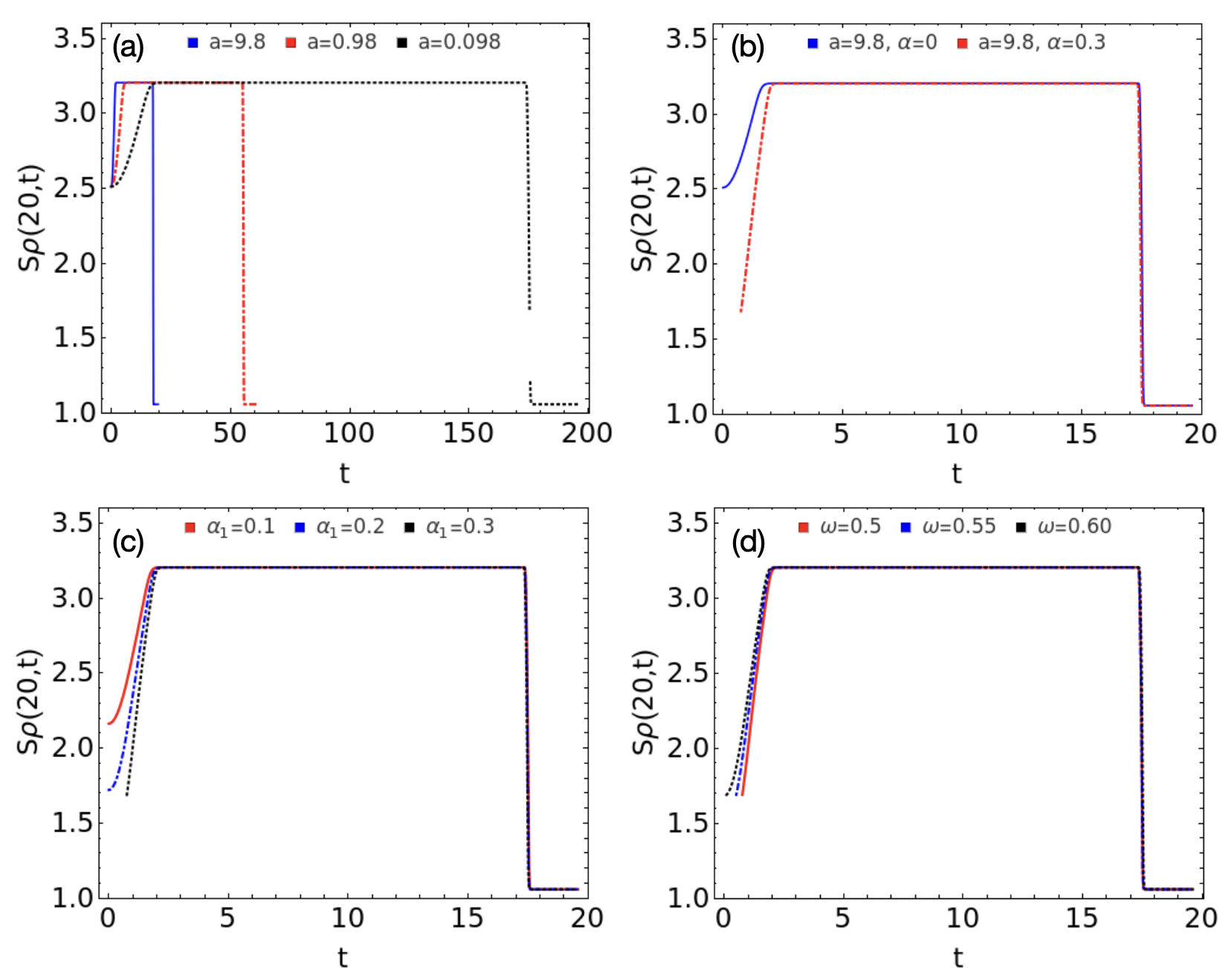}
    \caption{ Shannon entropy \( S_\rho(t) \) depicting the information-theoretic evolution of quantum droplets (QDs) at a fixed position \( x = 20 \), under a gravitational-like potential \( V(x,t) = -a(1 + \alpha \cos(\omega t))x \). 
(a) Comparison for constant \( a = 9.8 \) (blue), \( 0.98 \) (red dot-dashed), and \( 0.098 \) (black dashed) with \( \alpha = 0 \);
(b) Effect of modulation at \( a = 9.8 \): \( \alpha = 0 \) (blue) vs. \( \alpha = 0.3 \) (red dashed);
(c) Varying \( \alpha = 0.1, 0.2, 0.3 \) (red, blue dot-dashed, black dashed) at \( a = 9.8 \), \( \omega = 0.5 \);
(d) Varying \( \omega = 0.5, 0.55, 0.6 \) (red, blue dot-dashed, black dashed) at \( a = 9.8 \), \( \alpha = 0.3 \). All plots use \( \mu = \mu_0 = -\frac{2}{9} \), \( G_1 = -1 \), \( G_2 = 0.9999 \), and spatial coordinates are scaled by the harmonic oscillator length.}\label{fig9}
\end{figure}

Figure~(\ref{fig9}) presents the temporal evolution of Shannon entropy \( S_\rho(t) \) at a fixed position \( x = 20 \) under a gravitational-like potential of the form \( V(x,t) = -a(1 + \alpha \cos(\omega t))x \). In panel (a), as the strength of \( a \) increases from 0.098 (black dashed) to 9.8 (blue), the entropy rapidly rises to its peak value and converges towards \( S_\rho \approx 3.2 \) within a short timescale, indicating faster loss of coherence and higher phase-space delocalization for stronger gravity. This highlights the dominant role of \( a \) in accelerating the growth of entropy and improving phase mixing. In panel (b), the effect of temporal modulation is examined at a fixed \( a = 9.8 \), with \(\alpha = 0\) (blue) and \( \alpha = 0.3 \) (red dashed). Although both cases exhibit similar long-term behavior, the modulated system shows a noticeably higher entropy in the early stage \( t < 1 \), implying enhanced initial phase-space delocalization induced by periodic driving. Panels (c) and (d) explore the influence of varying modulation amplitude \( \alpha \) and frequency \( \omega \), respectively. At early times, different values of \( \alpha = 0.1, 0.2, 0.3 \) (panel c) and \( \omega = 0.5, 0.55, 0.6 \) (panel d) produce distinguishable slopes in entropy growth, reflecting sensitivity to drive parameters. However, as time progresses, the entropy values stabilize around \( S_\rho \approx 3.2 \) with a similar decay rate, suggesting that despite initial differences, the long-term coherence loss dynamics is robust against moderate variations in driving parameters. These findings reinforce the utility of Shannon entropy in quantifying the interplay of gravity and temporal modulation on the phase coherence of quantum droplets.

\begin{figure}
    \centering
    \includegraphics[width=1.0\linewidth]{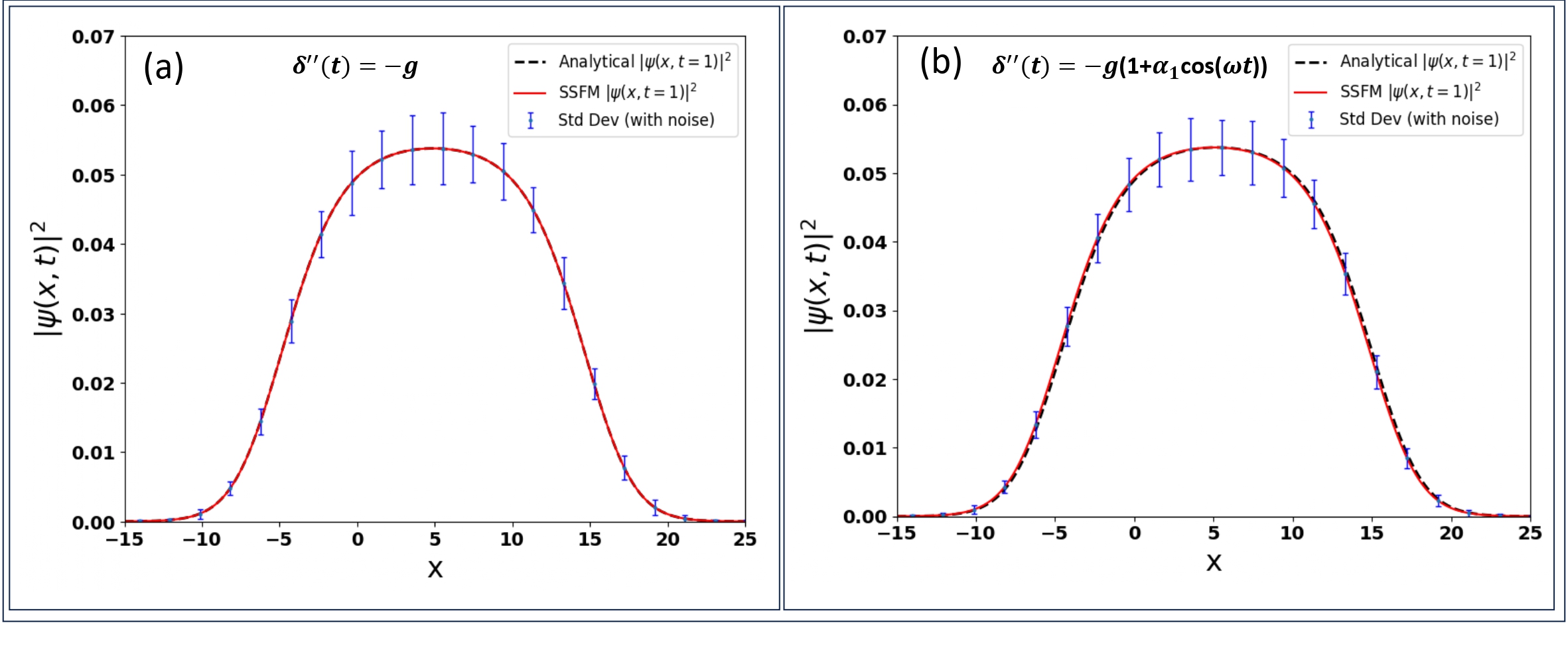}
    \caption{Stability analysis of obtained wavefunction solutions for both cases are performed by using the SSFM with \( dt = 0.0001 \), \( dx = 0.0488 \), and \( 10{,}000 \) iterations. (a) shows the evolution under a time-independent linear potential \( \gamma''(t) = -a \), while (b) represents a modulated potential \( \gamma''(t) = -a(1 + \alpha \cos(\omega t)) \). Analytical and numerical results are compared to assess QD stability. Simulation parameters: \( \mu = \mu_0 = -\frac{2}{9} \), \( G_1 = -1 \), \( G_2 = 0.9999 \), \( g = 9.8 \), \( N = 1 \), \( \alpha_1 = 0.3 \), and \( \omega = 0.5 \). }   \label{fig10}
\end{figure}

\section{Stability analysis} In this section, we assess the structural stability of 1D QDs by comparing exact analytical solutions of the 1D eGPE (equation \ref{eq:QD3}) with numerical simulations. The analytical solutions for both time-independent and time-dependent gravitational-like potentials with suitable choices of $\gamma(t)$ are used as initial states, and the system is evolved numerically to test their robustness. Numerical simulations are performed using the Split-Step Fourier Method (SSFM), a widely adopted pseudo-spectral technique for solving nonlinear Schrödinger-type equations~\cite{Muruganandam_CPC}. The evolution is carried out over 10,000 time steps with a temporal resolution of \( dt = 0.0001 \) and spatial resolution \( dx = 0.0488 \). Figure~\ref{fig10}(a) illustrates the QD dynamics under a time-independent gravitational potential with \( a(t) = -a \), whereas figure~\ref{fig10}(b) corresponds to the time-dependent potential \( a(t) = -a(1 + \alpha \cos(\omega t)) \), evaluated at time \( t = 1 \).

To probe the stability under perturbations, we introduce a weak stochastic noise of amplitude 1\% relative to the peak analytical density. The perturbed wavefunction is defined as: $\psi_{\text{noisy}}(x,0) = \psi(x,0) + R_w,$ where \( R_w \) denotes random white noise with an amplitude of 1\%. This perturbed state is then evolved numerically using the same 1D eGPE framework. The deviations between the analytical and noisy evolutions are quantified via standard deviation (Sd) at each spatial point and are indicated using vertical error bars in figure~(\ref{fig10}). The simulations reveal that the maximum deviation remains below 10\% for the time-independent case and below 12\% for the time-dependent scenario, confirming the stability of the analytical solutions under realistic perturbations. These results validate the robustness of self-bound QDs in gravitational-like potentials and support the applicability of these wavefunctions in modeling experimentally realizable ultradilute quantum fluids.

\section{Conclusion} In this study, we have comprehensively investigated the spatio- and temporal dynamics of 1D QDs under constant and time-dependent linear gravitational-like potentials. The system, modeled as a symmetric binary BEC and is governed by the 1D eGPE incorporating both repulsive cubic EMF and attractive quadratic BMF interactions. By employing an exact ansatz solution, we established consistency conditions linking the external modulation parameters to droplet dynamics. Our analysis first focused on a time-independent gravitational-like potential, where we observed that the QD’s falling velocity depends exclusively on the gravitational strength \( a \), independent of atom number or EMF nonlinearity- exhibiting classical Newtonian-like free-fall behavior. Interestingly, increasing the atom number primarily affects the density magnitude but not the velocity, indicating that gravitational acceleration governs the center-of-mass motion, while interatomic interactions shape internal structure. Upon introducing time-dependent modulations of the potential in the form \( V(x,t) = -a(1 + \alpha \cos(\omega t))x \), we demonstrated that both amplitude \( \alpha \) and frequency \( \omega \) significantly influence the droplet's trajectory, coherence, and localization properties. This control over dynamic behavior was further quantified using Shannon entropy and Wigner phase-space analyses. Furthermore, the analytical solutions were validated against full numerical simulations performed using the SSFM. Even under added perturbations, the QDs retained their structural integrity and coherence, confirming the stability and robustness of the derived solutions across both static and driven gravitational scenarios. These findings have significant implications for simulating gravitational effects and microgravity analogs in ultracold atomic systems. The demonstrated tunability of QD dynamics via external driving protocols can be harnessed for quantum sensing, interferometry, and precision gravimetry applications. Looking ahead, future investigations may include exploring multi-component QDs, higher-dimensional geometries, disordered or anharmonic potentials, and incorporating open-system dynamics to probe decoherence, dissipation, and quantum thermometric applications.

\end{document}